\documentclass[aps,preprint,showpacs,preprintnumbers,amsmath,amssymb,floatfix]{revtex4}
\usepackage{graphicx}
\begin{document}
\def\brho {\mbox{\boldmath $\rho$}}
\def\r {{\bf r}}

\title{
Formation and dynamics of many-boson fragmented states in 
attractive one-dimensional ultra-cold gases
}

\author{Alexej I. Streltsov, Ofir E. Alon and Lorenz S. Cederbaum}

\affiliation{Theoretische Chemie, Physikalisch-Chemisches Institut, Universit\"at Heidelberg,\\
Im Neuenheimer Feld 229, D-69120 Heidelberg, Germany}
\begin{abstract}
Dynamics of attractive ultra-cold bosonic clouds in one dimension are studied by solving the 
many-particle time-dependent Schr\"odinger equation.
The initially coherent wave-packet 
can dynamically dissociate into two parts when its energy exceeds a threshold value. 
Noticeably, the time-dependent Gross-Pitaevskii theory applied to the same initial state
does not show up the splitting. We call the split object {\it fragmenton}. 
It possesses remarkable properties: 
(1) it is two-fold fragmented, i.e., not coherent;
(2) it is dynamically stable, i.e., it propagates almost without dispersion; 
(3) it is delocalized, i.e., the two dissociated parts still communicate with one another.
A simple static model predicts the existence of fragmented states 
which are responsible for formation and dynamics of fragmentons.
\end{abstract}

\maketitle
Dynamics of attractive low-dimensional dilute Bose gases have attracted much
attention \cite{BS1,BS2,e1,e2,t1,t2,t3}. The formal equivalence of the Gross-Pitaevskii (GP) equation
used to describe these quantum gases and the non-linear Schr\"odinger equation
used in non-linear optics has led to transfer of knowhow and, most
importantly, to the prediction of bright solitons in attractive Bose gases \cite{BS1,BS2}. 
More recently, fascinating and challenging experiments on bright
matter-wave solitons have been reported \cite{e1,e2} and explained using GP theory,
making attractive Bose gases a cornerstone of GP theory. Nowadays, the GP
theory is widely utilized to study attractive dilute Bose gases.

GP theory is a mean-field approximation to the quantum {\it many-boson} problem,
in which the Bose system is assumed to remain {\it condensed} during its evolution in time.
Recently, the general validity of this assumption has been questioned 
in the theoretical study on dynamical depletion of attractive one-dimensional (1D) Bose systems 
at non-zero temperatures \cite{PRLB}.
In the present Letter we consider the dynamics of attractive 1D Bose systems 
from the full many-body point of view and relate the many-body and GP results.
We shall see that the dynamics of attractive 1D Bose gases 
can be much richer than anticipated before because along with condensed states 
the system supports a new kind of low-lying excited states which are {\it fragmented},
and thereby not available within the framework of GP theory. 
These low-lying fragmented excited states are responsible 
for the formation and evolution of dynamically-stable 
{\it fragmented} objects in attractive 1D Bose systems.
These objects are distinct from the solitons discussed above.

We solve the time-dependent many-body Schr\"odinger equation for N attractive bosons in 1D
interacting via contact potential $U(x-x')=\lambda_0\delta(x-x')$ 
in free space within the framework of the recently developed 
multi-configurational time-dependent Hartree for bosons (MCTDHB) \cite{MCTDHB}.
For comparison we also integrate the respective time-dependent GP equation. 
In GP theory the wavefunction has the form
$\Psi(x_1,\ldots,x_N,t) = \phi_1(x_1,{t})\phi_1(x_2,{t})\cdots\phi_1(x_N,{t})$, i.e., all
bosons reside in a single orbital $\phi_1(x,{t})$.
MCTDHB(M) goes much beyond GP in that the bosons are distributed over $M$ orbitals $\{\phi_i(x,t)\}$ and all
possible distributions are considered. The MCTDHB(M) ansatz reads 
$
\Psi(x_1,x_2,\ldots,x_N,t) = \sum_{\vec{m}} C_{\vec{m}}(t) 
\hat{\cal S}\phi_1(x_1,{{t}})\cdots\phi_2(x_{m_1+1},{t})
\cdots\phi_3(x_{m_1+m_2+1},{t})
\cdots\phi_M(x_{N},{t})
$
where $\hat{\cal S}$ is symmetrization operator and $\vec{m}=(m_1,m_2,\cdots m_M)$ occupation numbers.
Within this theory a quantitative description of the time evolution of bosonic systems is achieved by optimizing
all $M$ orbitals used to construct the many-body expansion and the expansion coefficients themselves 
at each point in time utilizing time-dependent variational principle \cite{MCTDHB}.

At t=0 we take as initial condition a {\em totally condensed} atomic cloud of a very-well defined shape and  
propagate the many-body MCTDHB(2) and GP equations in time and compare the obtained densities.
In this work we take initial coherent wave-packets where all bosons are in either $Sech \left[ \gamma x \right]$- or Gaussian-shaped 
normalized functions. These choices represent different ways to generate the initial conditions,
leading, however, to similar physical results.
Varying the widths of the initial wave packet and keeping the number of bosons and their interparticle 
interaction strength fixed allows us to change the energy of the initial wave packet and to activate thereby 
different excited states of the many-boson attractive system. In this study we allow only for 
initial wave-packets which energetically do not exceed the formation energy of two-hump soliton trains.
Thus, {\em by construction} all the results observed in the present study cannot be attributed to soliton trains.

\begin{figure}
\includegraphics[width=8.6cm, angle=-0]{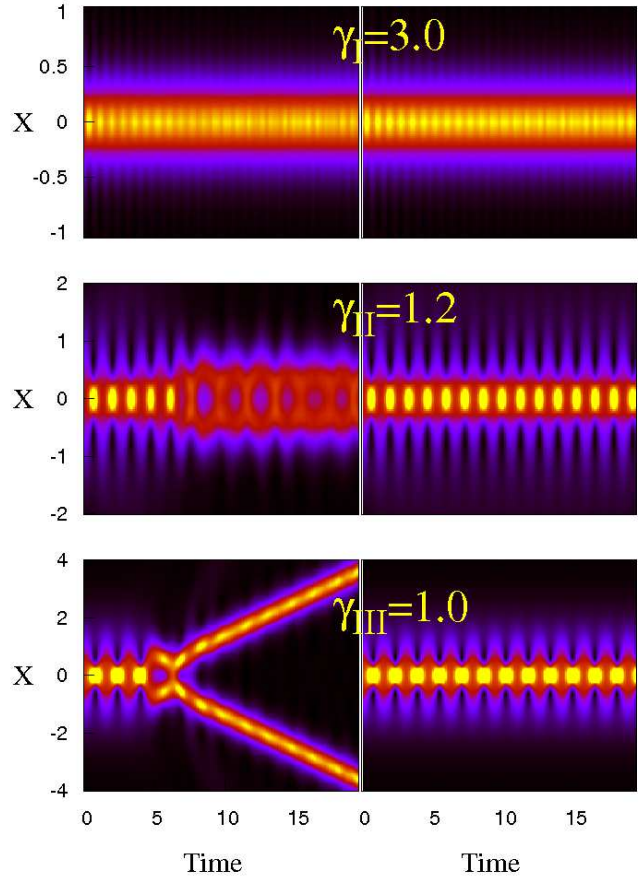}
\caption {(Color online) MCTDHB many-body (left) vs. GP (right) dynamics for N=1000 
attractive ($\lambda_0=-0.008$) bosons in one-dimension.
The densities are plotted as a function of time for
initially-coherent wave-packets $Sech \left[ \gamma x \right]$ of different widths.
While the GP dynamics shows "breathing" oscillations for all initial widths, the many-body dynamics changes 
dramatically from "breathing" oscillations ($\gamma_{I}=3.0$), to attempts for dynamical splitting  ($\gamma_{II}=1.2$),
and to dynamical dissociation into two parts ($\gamma_{III}=1.0$).
All quantities  are dimensionless.
}
\label{fig1}
\end{figure}

In Fig.~\ref{fig1} we present the results of the many-body Schr\"odinger equation 
for N=1000 attractive bosons with $\lambda_0=-0.008$ for several {\em Sech}-shaped initial wave-packets of different widths.
Here and in the following we work in dimensionless units which are readily arrived at by
introducing a convenient length scale $L$ (say, the scattering length) and dividing
the Hamiltonian by $\frac{\hbar^2}{m L^2}$, where $m$ is the mass of a boson.
The densities are plotted as a function of time.
The many-body MCTDHB densities depicted in the left column can be compared with the respective time-dependent GP densities 
plotted in the right column obtained for the same system with identical initial conditions.
The ground eigenstate of the many-boson attractive system in free space at the GP level of description 
is known analytically -- it is the famous $Sech \left[ \gamma x \right] $ 
with optimal exponent $\gamma=\frac{|\lambda_0|(N-1)}{2}$. For the system under investigation the optimal exponent  
corresponding to the ground eigenstate is $\gamma \approx4.0$.
In the first study we take a slightly broader initial {\em Sech}-function with $\gamma_{I}=3.0$. 
The corresponding many-body and GP dynamics are depicted in the upper left and right panels of Fig.~\ref{fig1}, respectively.
We see that the width of the wave-packet changes in time rather periodically 
describing thereby "breathing" of the attractive atomic cloud. 
The many-body and GP dynamics are quite similar (but not identical) 
confirming the solitonic character of the ground state also at the many-body level of description. 
In the second study presented in the middle panels of Fig.~\ref{fig1}
we compare many-body and GP dynamics for a broader initial {\em Sech}-function with $\gamma_{II}=1.2$.
We see that dynamics of the first five "breathing" oscillations are quite similar within both theories,
while at longer times the many-body theory shows attempts of the density to split into two parts.
The bottom panels of Fig.~\ref{fig1} show the results of the third study, 
where we compare many-body and GP propagations of an even broader initial {\em Sech}-function with $\gamma_{III}=1.0$.
Again, the first three "breathing" oscillations of the GP dynamics resemble the respective many-body ones, then, however,
the many-body dynamics reveal a fascinating feature of the evolution 
-- the initially condensed single {\em Sech}-shaped cloud dissociates into two smaller equal clouds which move 
symmetrically apart from each other with approximately constant velocity. 

We stress several points.
First, the dynamical many-body splitting is a general phenomenon found for many initial conditions and 
for wave-packets of other shapes as well, e.g., for symmetric and asymmetric Gaussian-shaped profiles. 
Asymmetric initial packets dissociate into unequal parts.
Second, the phenomenon takes place only if the energy of the initial cloud exceeds some threshold value (see below).
Third, by increasing the widths of the initial wave packets we increase their total energies 
making thereby different excited states of the quantum many-boson system energetically accessible.
The existence of the splitting phenomenon found only at the many-body level of description indicates that
in attractive 1D Bose systems there is a new kind of low-lying excited states, 
which is not available within the framework of GP theory.

\begin{figure}[ht]
\includegraphics[width=8.6cm,angle=-90]{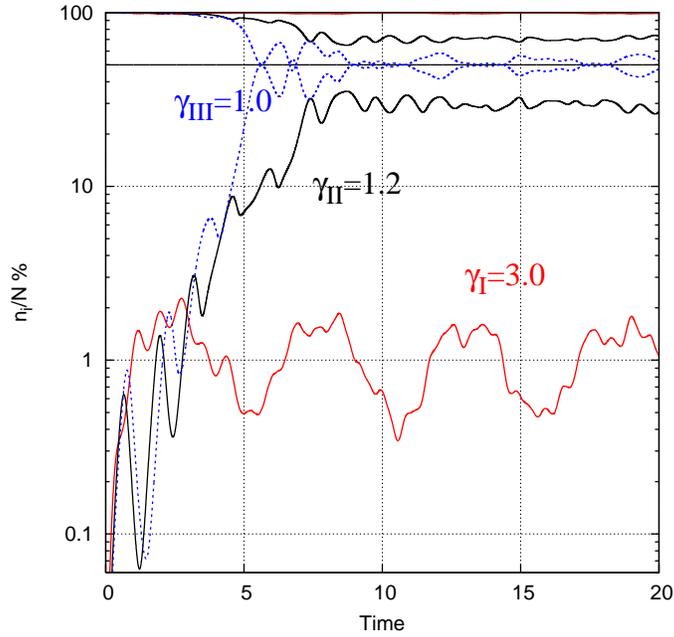}
\caption [kdv]{(Color online) 
Evolutions of natural occupation numbers $n_1/N$ and $n_2/N$ of the reduced one-body densities 
plotted in the left panels of Fig.~\ref{fig1} on a log scale in \%. At t=0 all initial states are condensed: $n_1/N=100 \%$.
In the first study ($\gamma_{I}=3.0$) the dynamics describe "breathing" oscillations only.
In the second ($\gamma_{II}=1.2$) and third ($\gamma_{III}=1.0$) studies the systems evolve to be
two-fold fragmented after few "breathing" oscillations. 
Notice that in GP $n_1/N$ is always 100 \%.
All quantities are dimensionless. 
}
\label{fig2}
\end{figure}

To get a deeper insight into the physics, we investigate next the many-body structures of the evolving wave-packets
in some details. By diagonalizing the reduced one-body density matrix at each time step we observe how its eigenvalues, i.e., 
natural occupation numbers ($n_i$) and its eigenvectors, i.e., natural orbitals evolve during the dynamics.
The evolutions of the natural occupation numbers for all the three initial conditions reported in Fig.~\ref{fig1} 
are plotted in Fig.~\ref{fig2}. We choose log scale to plot the time evolution of the $n_i/N$ in \%. 
At t=0 all the bosons of the initial wave-packets reside in one and the same natural orbital indicating that  
these states, according to the usual definition \cite{Penrose} are condensed. 
In time we see that another orbital acquires some population revealing the depletion of the initially-condensed state.
Interestingly, during the times where the system shows "breathing" dynamics, 
the changes of the respective natural occupations are moderate.
Indeed, for $\gamma_{I}=3.0$ where GP theory predicts very similar breathing dynamics,
the occupation of the second natural orbital does not exceed 2\%.
In the other studies ($\gamma_{II}=1.2$,$\gamma_{III}=1.0$) the second natural occupation number oscillatory increases until 10-15\%
as long as the density exhibits "breathing" oscillations.  
Then, the system fragments (see Fig.~\ref{fig2}).

Let us concentrate on the third study. As time proceeds, the occupation of the second natural orbital increases
and around the point where the cloud starts to dissociate into two parts, the two natural occupation numbers become equal and macroscopic.
According to the usual definition \cite{Nozieres} such a system is two-fold fragmented.
Clearly, GP theory is inapplicable here.
Further evolution reveals oscillations of the occupation numbers around this ideally two-fold fragmented state.
This results from the fact that the respective natural orbitals themselves are delocalized objects,
so that even at very large separations the dissociated parts remain connected and can "talk" to each other (see below).
The two separated packets propagate almost without dispersion like ordinary solitons (see also Fig.~\ref{fig1}). 
We call this new physical object {\it fragmenton} because it
combines macroscopic fragmentation of the wave function and dynamical properties of a soliton. 
We recall that in contrast to the ordinary soliton or soliton train solutions which are {\it coherent} objects,
fragmenton is fragmented, i.e., characterized by macroscopic occupations of several (here two) natural orbitals. 

\begin{figure}[ht]
\includegraphics[width=8.6cm,angle=-0]{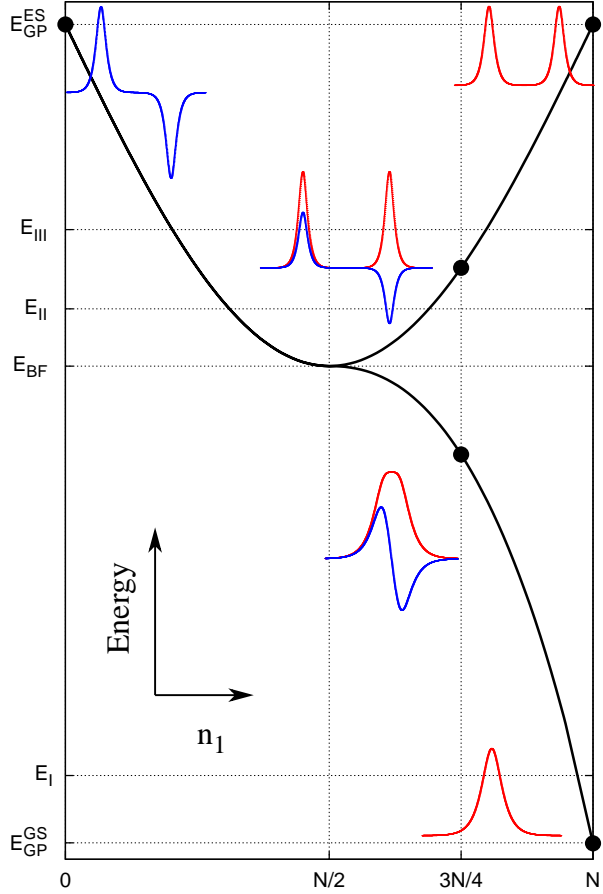}
\caption [kdv] {(Color online)
Energy diagram $E(n_1)$ of the model system introduced in Eqs.~(\ref{ansatz},\ref{BMF_energy})
for two-fold fragmented $|n_1,n_2\rangle$ states $n_1+n_2=N$.
There are two branches of solutions bifurcating at $n_1=N/2$ (bifurcation energy $E_{BF}$).
Shown are also the orbitals $\sqrt{n_i/N} \phi_i$ corresponding to 
$n_1=3N/4$ at both branches and to the GP solutions of $n_1=N$ and $n_2=N$ at
energies $E^{GS}_{GP}$ and $E^{ES}_{GP}$.
$E_{I},E_{II},E_{III}$ are the energies of the three cases studied in Fig.~\ref{fig1}. 
To activate a fragmenton the energy of the initial cloud must exceed $E_{BF}$.
All quantities are dimensionless.
}
\label{fig3}
\end{figure}

For the sake of interpretation, we construct a simple static model 
which is able to describe two-fold fragmented states $|n_1,n_2\rangle$,
where $n_1$ bosons reside in $\phi_1$ and $n_2=N-n_1$ in $\phi_2$ orbitals.
As the orbitals constituting this state we choose delocalized gerade and ungerade superpositions 
of two identical {\em Sech}-functions with equal exponent $\gamma_m$ placed at $\pm X_0$ with respect to the origin:
\begin{equation}
\phi_{1,2}=Y_{1,2}\left[Sech(\gamma_m (x-X_0)) \pm Sech(\gamma_m (x+X_0)) \right],
\label{ansatz}
\end{equation}
where $Y_{1,2}$ are the respective normalization factors. 
Our goal is to find within this ansatz the lowest energy eigenstates of the N identical interacting bosons.
We consider the exponent $\gamma_m$ and position $X_0$ of the {\em Sech}-functions as variational parameters.
Additionally we can change the occupation number $n_1$ ($n_2=N-n_1$). 
The expectation energy of a two-fold fragmented state is known \cite{BMF}:
\begin{eqnarray}
& & E(n_1)= n_1 h_{11} +  \lambda_0 \frac{n_1(n_1-1)}{2}\int|\phi_1|^4 d{x}+ n_2 h_{22}  \nonumber \\ 
      & & \qquad \qquad + \lambda_0 \frac{n_2(n_2-1)}{2}\int|\phi_2|^4 d{x}+  
        2 \lambda_0 n_1 n_2 \int|\phi_1|^2 |\phi_2|^2 d{x} 
\label{BMF_energy}
\end{eqnarray}
where $h_{ii}= \langle \phi_i|-\frac{1}{2}\frac{d^2}{d x^2}|\phi_i \rangle$ are the expectation values of the kinetic energy operator. 
The minima of this energy functional are plotted in Fig.~\ref{fig3} for our system of N=1000 bosons with $\lambda_0=-0.008$
as a function of $n_1$.

The ground state corresponds to the situation where all the bosons are condensed in one localized orbital $|N,0\rangle$.
For an open system this is {\em exactly} the bright soliton GP solution with $X_0=0$ and $\gamma_m\equiv\frac{|\lambda_0|(N-1)}{2}$,
which is reproduced by our model. In the right lower corner of Fig.~\ref{fig3} we schematically plot this function. 
If we increase the population of the second orbital, the total energy of the system increases
and the density starts to split, i.e., $X_0$ has non-zero values.
For example for $n_1=3N/4$, the minimum energy $E/N=-1.7135$ is obtained 
for individual {\em Sech}-functions slightly ($X_0=0.2907$) separated from each other and have exponent $\gamma_m=2.9839$.
In Fig.~\ref{fig3} we depict this energy and respective orbitals $\sqrt{n_i/N} \phi_i$.
However, our simple model accounts for two branches of physically different solutions.
For any given $n_1$ --  there is another branch of solutions which is much higher in energy 
and characterized by large values of optimal $X_0$, i.e., by two very-well separated humps. 
Indeed, for $n_1=3N/4$ such a solution has optimal energy $E/N=-1.2586$ with exponent $\gamma_m=2.7480$ and separation $X_0=4.0$.
We depict the respective energy point and pair of orbitals in Fig.~\ref{fig3}.
Reducing $n_1$ further, we increase the energy of the lower branch and decrease the energy of the upper branch
which approaches its minimum at $n_1=n_2=N/2$. At this point, as we can see in Fig.~\ref{fig3},
one encounters a bifurcation of the branches. Increasing the occupation of the second orbital further, 
the energy increases until at $n_1=0,n_2=N$  we arrive at the two-soliton anti-phase (ungerade) solution,
plotted in the left upper corner of Fig.~\ref{fig3}.

There are three points of relevance indicated in Fig.~\ref{fig3}:
Ground state (GS), excited state (ES) and the bifurcation point (BF) which corresponds to the ideally fragmented ($n_1=n_2=N/2$) 
state of minimal energy. 
In our model the optimal energies of all these states have a unique form $E_{state}=-N{\gamma_{state}}^2/6$, 
with different optimal exponents $\gamma_{state}$. 
Indeed, the ground state energy of the bright soliton is known: $E^{GS}_{GP}=-N{\gamma_{GS}}^2/6$ with $\gamma_{GS}=|\lambda_0|(N-1)/2$.
The energies of the coherent two-soliton gerade and ungerade states can be obtained 
by minimizing the GP energy functional assuming zero overlap of the constituting solitons. Then these degenerate 
energies are $E^{ES}_{GP}=-N{\gamma_{ES}}^2/6$ with exponent $\gamma_{ES}=|\lambda_0|(N-1)/4$.
Finally, assuming that the {\em Sech}-functions forming the orbitals of ideally fragmented state
do not overlap,  we get by minimizing Eq.~(\ref{BMF_energy}) the optimal exponent 
$\gamma_{BF}=|\lambda_0|(3 N/2-1)/4$ and bifurcation energy $E_{BF}=-N{\gamma_{BF}}^2/6$.
From these energies we see that the ideally two-fold fragmented state is separated from the ground state by the energy gap
$\Delta E_{FR}\approx7 N^3 {\lambda_0}^2 /384$. This energy can be viewed as the threshold for the activation of the fragmenton.
The energy difference between the ground and excited GP states is $\Delta E_{ES}\approx 12 N^3 {\lambda_0}^2 /384$.
Therefore, in attractive one-dimensional Bose gases two-fold fragmented delocalized states are always energetically more favorable  
than the respective two-soliton coherent gerade and ungerade states. 

Now we can interpret the wave-packet evolutions presented in Fig.~\ref{fig1} in terms of the states of the system (see Fig.~\ref{fig3}).
If the energy of the initial wave-packet relative to the ground state does not exceed 
the fragmenton activation energy $\Delta E_{FR}$, the model permits dynamics only within the states of the lowest branch. 
Indeed, in the first study the energy $E_{I}$ of the initial wave packet (see Fig.~\ref{fig3})
is insufficient for activating the fragmenton. 
The many-body scenario plotted in the left upper panel of Fig.~\ref{fig1} shows  small-amplitude "breathing" oscillations 
around the ground state.  In contrast, if the activation threshold is overcome, as it happens in the second ($E_{II}$) 
and in particular in the third ($E_{III}$) study, the fragmenton channel becomes accessible energetically 
and strongly affects the dynamics.
This is reflected in the left middle and lower panels of Fig.~\ref{fig1} where the initial wave packet
attempts to split and successfully dissociates into two parts, respectively.
The respective $E_{I}$, $E_{II}$ and $E_{III}$ energies are shown in Fig.~\ref{fig3}.

Let us summarize. 
We explore the challenging many-body dynamics of attractive ultra-cold bosonic clouds in 1D
by varying the width and thereby the total energy of the initial coherent {\em Sech}-shaped wave-packet. 
We solve the many-boson time-dependent Schr\"odinger equation and show that the many-body dynamics of attractive condensates
can differ considerably from the respective GP dynamics. We find that when the initial energy of the condensate 
exceeds some threshold, the condensed cloud at the many-body level of description tends to split and can dynamically 
dissociate into two separate parts, while at the GP level it reveals only "breathing" dynamics and remains unsplit. 
To understand this phenomenon we first analyze the many-body wave-function 
and attribute the split clouds to a two-fold fragmented delocalized state. 
We call this new class of dynamical solutions {\it fragmentons}
because on the one hand they propagate almost without dispersion, and on the other hand, 
in contrast to solitons, they are not coherent objects.
A simple static model is introduced to show that the
two-fold fragmented state constructed from in- and anti- phase two-soliton-like functions  
is energetically much more favorable than coherent two-soliton solutions.
To estimate the window for fragmenton formation we provide simple estimations of the energies of the involved states 
in terms of the number of particles $N$ and their interaction strength $\lambda_0$.

Our findings imply that the dynamics of attractive Bose gases in low dimensions is much richer than anticipated before
and we hope that our work will stimulate experiments on fragmentons.

\begin{acknowledgments}
Financial support by DFG is acknowledged.
\end{acknowledgments}
 
\end{document}